\documentclass[conference]{IEEEtran}
\usepackage{amsmath,amssymb,amsfonts}
\usepackage{algorithmic}
\usepackage{graphicx}
\usepackage{subcaption}
\usepackage{textcomp}
\usepackage{xcolor}

\usepackage{url}
\usepackage{cleveref}
\usepackage{todonotes}
\usepackage{csquotes}
\usepackage{blindtext}

\usepackage[style=ieee,backend=biber,bibencoding=utf8]{biblatex}

\graphicspath{{./img/}}

\def\BibTeX{{\rm B\kern-.05em{\sc i\kern-.025em b}\kern-.08em
		T\kern-.1667em\lower.7ex\hbox{E}\kern-.125emX}}

\newcommand{\projectLegacyName}{\textit{ExplorViz Legacy}}

\newcommand{\projectName}{\textit{ExplorViz}}

\begin{document}
\pagenumbering{gobble}
\title{Modularization of Research Software for Collaborative Open Source Development}

\author{\IEEEauthorblockN{Christian Zirkelbach}
\IEEEauthorblockA{\textit{Software Engineering Group} \\
\textit{Kiel University}\\
Kiel, Germany \\
email: czi@informatik.uni-kiel.de}
\and
\IEEEauthorblockN{Alexander Krause}
\IEEEauthorblockA{\textit{Software Engineering Group} \\
\textit{Kiel University}\\
Kiel, Germany \\
email: akr@informatik.uni-kiel.de}
\and
\IEEEauthorblockN{Wilhelm Hasselbring}
\IEEEauthorblockA{\textit{Software Engineering Group} \\
\textit{Kiel University}\\
Kiel, Germany \\
email: wha@informatik.uni-kiel.de}
}

\maketitle

\begin{abstract}
Software systems evolve over their lifetime.
Changing conditions, such as requirements or customer requests make it inevitable for developers to perform adjustments to the underlying code base.
Especially in the context of open source software where everybody can contribute, requirements can change over time and new user groups may be addressed.
In particular, research software is often not structured with a maintainable and extensible architecture.
In combination with obsolescent technologies, this is a challenging task for new developers, especially, when students are involved.
In this paper, we report on the modularization process and architecture of our open source research project \projectName\ towards a microservice architecture.
The new architecture facilitates a collaborative development process for both researchers and students.
We describe the modularization measures and present how we solved occurring issues and enhanced our development process.
Afterwards, we illustrate our modularization approach with our modernized, extensible software system architecture and highlight the improved collaborative development process.
Finally, we present a proof-of-concept implementation featuring several developed extensions in terms of architecture and extensibility.
\end{abstract}


\section{Introduction}\label{sec-introduction}

Software systems are continuously evolving during their lifetime.
Changing contexts, legal, or requirement changes such as customer requests make it inevitable for developers to perform modifications of existing software systems.
Open source software is based on the open source model, which addresses a decentralized and collaborative software development.
Open research software~\cite{ResearchSoftware:Goble2014} is available to the public and enables anyone to copy, modify, and redistribute the underlying source code.
In this context, where anyone can contribute code or feature requests, requirements can change over time and new user groups may appear.
Although this development approach features a lot of collaboration and freedom, the resulting software does not necessarily constitute a maintainable and extensible underlying architecture.
Additionally, employed technologies and frameworks can become obsolescent or are not updated anymore.
In particular, research software is often not structured with a maintainable and extensible architecture~\cite{CiSE2018}.
This causes a challenging task for developers during the development, especially when inexperienced collaborators like students are involved.
Based on several drivers, like technical issues or occurring organization problems, many research and industrial projects need to move their applications to other programming languages, frameworks, or even architectures.
Currently, a tremendous movement in research and industry constitutes a migration or even modernization towards a microservice architecture, caused by promised benefits like scalability, agility, and reliability~\cite{OttoMicroservices:2017}.
Unfortunately, the process of moving towards a microservice-based architecture is difficult, because there a several challenges to address from both
technical and organizational perspectives~\cite{MigrationToMicroservicesSurvey:2018}.
In this paper, we report on the modularization process of our open source research project~\projectName\ towards a more collaboration-oriented development process on the basis of a microservice architecture.
We later call the outdated version~\projectLegacyName, and the new version just~\projectName.

The remainder of this paper is organized as follows.
In~\Cref{sec-problemStatement}, we illustrate our problems and drivers for a modularization and architectural modernization.
Afterwards, we illustrate our software system and underlying architecture of~\projectLegacyName~in~\Cref{sec-legacySystem}.
The following modularization and modernization process as well as the target architecture of~\projectName~are described in~\Cref{sec-modularizationProcess}.
\Cref{sec-proofOfConcept} introduces our proof of concept in detail, including an evaluation based on several developed extensions.
Our ongoing work in terms of achieving an entire microservice architecture is presented in~\Cref{sec-ongoingWork}.
\Cref{sec-relatedWork} discusses related work on modularization and modernization towards microservice architectures.
Finally, the conclusions are drawn and an outlook is given.
\section{Problem Statement}\label{sec-problemStatement}

The open source research project~\projectName~started in 2012 as part of a PhD thesis and is further developed and maintained until today.
\projectName~enables a live monitoring and visualization of large software landscapes~\cite{ExplorViz,ECIS2015}.
The tool has the objective to aid the process of system and program comprehension for developers and operators.
We successfully employed the software in several collaboration projects~\cite{ICSATutorial:2017,Elsevier2017} and experiments~\cite{VISSOFT2015hierarchical,ExplorVizControlledExperiment2015}.
The project is developed from the beginning on GitHub with a small set of core developers and many collaborators (more than 30 students) over the time.
Several extensions have been implemented since the first version, which enhanced the tool's feature set.
Unfortunately, this led to an unstructured architecture due to an unsuitable collaboration and integration process.
In combination with technical debt and issues of our employed software framework and underlying architecture, we had to perform a technical and process-oriented modularization. 
Since 2012, several researchers, student assistants, and a total of 25 student theses as well as multiple projects contributed to~\projectName.
We initially chose the Java-based Google Web Toolkit~(GWT)~\cite{Software:GWT}, which seemed to be a good fit in 2012, since Java is the most used language in our lectures.
GWT provides different wrappers for Hypertext Markup Language~(HTML) and compiles a set of Java classes to JavaScript~(JS) to enable the execution of applications in web browsers.
Employing GWT in our project resulted in a monolithic application (hereinafter referred to as \projectLegacyName), which introduced certain problems over the course of time.

\subsubsection{Extensibility \& Integrability}
\projectLegacyName's concerns are divided in core logic (core), e.g., predefined software visualizations, and extensions.
When~\projectLegacyName~was developed, students created new git branches to implement their given task, e.g., a new feature.
However, there was no extension mechanism that allowed the integration of features without rupturing the core's code base.
Therefore, most students created different, but necessary features in varying classes for the same functionality.
Furthermore, completely new technologies were utilized, which introduced new, sometimes even unnecessary (due to the lack of knowledge), dependencies.
Eventually, most of the developed features could not be easily integrated into the master branch and thus remained isolated in their feature branch.

\subsubsection{Code Quality \& Comprehensibility}
After a short period of time, modern JS web frameworks became increasingly mature.
Therefore, we started to use GWT's JavaScript Native Interface~(JSNI) to embed JS functionality in client-related Java methods.
Unfortunately, JSNI was overused and the result was a partitioning of the code base.
Developers were now starting to write Java source code, only to access JS, HTML, and Cascading Style Sheets~(CSS).
Furthermore, the integration of modern JS libraries in order to improve the user experience in the frontend was problematic.
Additionally, Google announced that JSNI would be removed with the upcoming release of Version 3, which required the migration of a majority of client-related code.
Google also released a new web development programming language, named \textit{DART}, which seemed to be the unofficial successor of GWT.
Thus, we identified a potential risk, if we would perform a version update.
Eventually, JSNI reduced our code quality.
Our remaining Java classes further suffered from ignoring some of the most common Java conventions and resulting bugs.
Students of our university know and use supporting software for code quality, e.g., static analysis tools such as \textit{Checkstyle}~\cite{Software:Checkstyle} or \textit{PMD}~\cite{Software:PMD}.
However, we did not define a common code style supported by these tools in~\projectLegacyName.
Therefore, a vast amount of extensions required a lot of refactoring, especially when we planned to integrate a feature into the core.

\subsubsection{Software Configuration \& Delivery}
In \textit{ExplorViz Legacy}, integrated features were deeply coupled with the core and could not be easily taken out.
Often, users did not need all features, but only a certain subset of the overall functionality.
Therefore, we introduced new branches with different configurations for several use cases, e.g., a live demo.
Afterwards, users could download resulting artifacts, but the maintenance of related branches was cumbersome.
Summarized, the stated problems worsened the extensibility, maintainability, and comprehension for developers of our software.
Therefore, we were in need of modularizing and modernizing \textit{ExplorViz}.
\section{\projectLegacyName}\label{sec-legacySystem}
\begin{figure*}[!ht]
	\centering
	\includegraphics[width=0.7\textwidth]{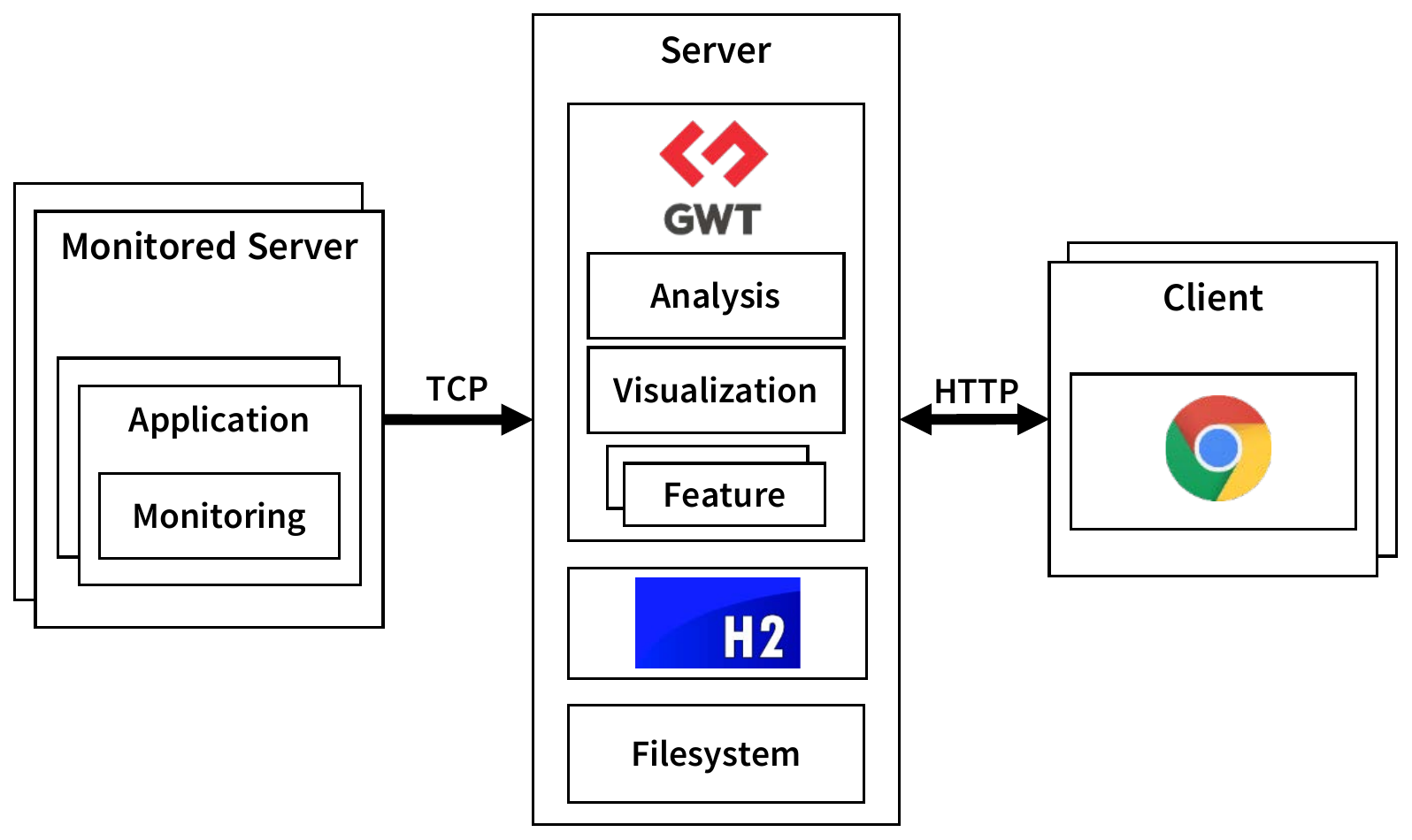}
	\caption{Architectural overview and software stack of~\projectLegacyName.}
	\label{fig:architecture-stack-before}
\end{figure*}
The overall architecture and the employed software stack of~\projectLegacyName~is shown in~\Cref{fig:architecture-stack-before}.
We are instrumenting applications, regardless whether they are native applications or deployed artifacts in an application server like Apache Tomcat.
The instrumentation is realized by our monitoring component, which employs in the case of Java \textit{AspectJ}, an aspect-oriented programming extension for Java~\cite{Software:AspectJ}.
\textit{AspectJ} allows us to intercept an application by bytecode-weaving in order to gather necessary monitoring information for analysis and visualization purposes.
Subsequently, this information is transported via (Transmission Control Protocol~(TCP) towards a server, which hosts our GWT application.
This part represents the two major components of our architecture, namely \textit{analysis} and \textit{visualization}.
The \textit{analysis} component receives the monitoring information and reconstructs traces.
These traces are stored in the file system and describe a software landscape consisting of monitored applications and communication in-between.
Our user-management employs a \textit{H2 database}~\cite{Software:H2} to store related data.
The software landscape \textit{visualization} is provided via Hypertext Transfer Protocol~(HTTP) and is accessible by clients with a web browser.
GWT is an open source framework, which allows to develop JS front-end applications in Java.
It facilitates the usage of Java code for server (backend) and client (frontend) logic in a single web project.
Client-related components are compiled to respective JS code.
The communication between frontend and backend is handled through asynchronous remote procedure calls based on HTTP.
In~\projectLegacyName, the advantages of GWT proved to be a drawback, because every change affects the whole project due to its single code base.
New developed features were hard-wired into the software system.
Thus, a feature could not be maintained, extended, or replaced by another component with reasonable effort.
This situation was a leading motivation for us to look for an up-to-date framework replacement.
We intended to take advantage of this situation and modularize our software system in order to move from a monolithic, to a distributed (web) application divided into separately maintainable and deployable backend and frontend components.

\section{Modularization Process and Architecture of~\projectName}\label{sec-modularizationProcess}
The previously mentioned drawbacks in~\projectLegacyName~and recent experience reports in literature about successful applications of alternative technologies, e.g., Representational State Transfer~(REST or RESTful) Application Programming Interfaces~(API)~\cite{SOAPtoREST,RESTfulWebServicesDevelopmentChecklist}, were triggers for a modularization and modernization.
In~\cite{EMLS:2018}, we gave a very brief description on the modernization process of \projectName~towards a microservice architecture.
During the modularization planning phase, we started with a requirement analysis for our modernized software system and identified technical and development process related impediments in the project.
We kept in mind that our focus was to provide a collaborative development process, which encourages developers to participate in our research project~\cite{EMLS:2018}.
Furthermore, developers, especially inexperienced ones, tend to have potential biases during the development of software, e.g., they make decisions on their existing knowledge instead of exploring unknown solutions~\cite{LitReviewDecisionMaking:2017}.
A more detailed description of decision triggers and the decision making process will be published in a technical report~\cite{ExtendedVersionCOLLA2019}.
In general, there exist many drivers and barriers for microservice adoption~\cite{EMISA2019}.
Typical barriers and challenges are the required additional governance of distributed, networked systems and the decentralized persistence of data.

As a result of this process, we agreed on building upon an architecture based on microservices as shown in~\Cref{fig:architecture-stack-after}.
This architectural style offers the ability do divide monolithic applications into small, lightweight, and independent services, which are also separately deployable~\cite{HolgerMicroservices:2018, OttoMicroservices:2017, MicroservicesEvolution2017, SMSMicroservices2016}.
However, the obtained benefits of a microservice architecture can bring along some drawbacks, such as increased overall complexity and data consistency~\cite{Carrasco:2018}.

\subsubsection{Extensibility \& Integrability}
In a first step, we modularized our GWT project into two separated projects, i.e., backend and frontend, which are now two self-contained microservices.
Thus, they can be developed technologically independent and deployed on different server nodes.
This allows us to exchange the microservices, as long as we take our specified APIs into account.
The backend is implemented as a Java-based web service based on the \textit{Jersey Project}~\cite{Software:Jersey}, which provides a RESTful API via HTTP for clients.
Furthermore, we replaced our custom-made monitoring component by the monitoring framework \textit{Kieker}~\cite{Kieker}.
This framework provides an extensible approach for monitoring and analyzing the runtime behavior of distributed software systems.
Monitored information is sent via TCP to our backend, which employs the filesystem and \textit{H2} database for storage.
The frontend uses the JS framework \textit{Ember.js}, which enables us to offer visualizations of software landscapes to clients with a web browser~\cite{Software:Ember}. 
Since \textit{Ember} is based on the model-view-viewmodel architectural pattern, developers do not need to manually access the Document Object Model and thus need to write less source code.
\textit{Ember} uses \textit{Node.js} as execution environment and emphasizes the use of components in web sites, i.e., self-contained, reusable, and exchangeable user interface fragments~\cite{Software:Node.js}. 
We build upon these components to encapsulate distinct visualization modes, especially for extensions.
Communication, like a request of a software landscape from the backend, is abstracted by so-called \textit{Ember} adapters.
These adapters make it easy to request or send data by using the convention-over-configuration pattern.
The introduced microservices, namely backend and frontend, represent the core of~\projectName.
As for future extensions, we implemented well-defined extension interfaces for both microservices, that allow their integration into the core.

\subsubsection{Code Quality \& Comprehensibility}
New project developers, e.g., students, do not have to understand the complete project from the beginning.
They can now extend the core by implementing new mechanics on the basis of a plug-in extension.
Extensions can access the core functionality only by a well-defined read-only API, which is implemented by the backend, respectively frontend.
This high level of encapsulation and modularization allows us to improve the project, while not breaking extension support.
Additionally, we do no longer have a conglomeration between backend and frontend source code, especially the mix of Java and JS, in single components.
This eased the development process and thus reduced the number of bugs, which previously occurred in~\projectLegacyName.
Another simplification was the use of \textit{json:api}~\cite{Software:json:api} as data exchange format specification between backend and frontend, which introduced a well-defined JavaScript Object Notation~(JSON) format with attributes and relations for data objects.
\subsection{Software Configuration \& Delivery}
One of our goals was the ability to easily exchange the microservices.
We fulfill this task by employing frameworks, which are exchangeable with respect to their language domain, i.e., Java and JS.
We anticipate that substituting these frameworks could be done with reasonable effort, if necessary.
Furthermore, we offer pre-configured artifacts of our software for several use cases by employing Docker images.
Thus, we are able to provide containers for the backend and frontend or special purposes, e.g., a fully functional live demo.
Additionally, we implemented the capability to plug-in developed extensions in the backend, by providing a package-scanning mechanism.
The mechanism scans a specific folder for compiled extensions and integrates them at runtime.

\section{Proof-of-Concept Implementation}\label{sec-proofOfConcept}

\begin{figure*}[!ht]
	\centering
	\includegraphics[width=\textwidth]{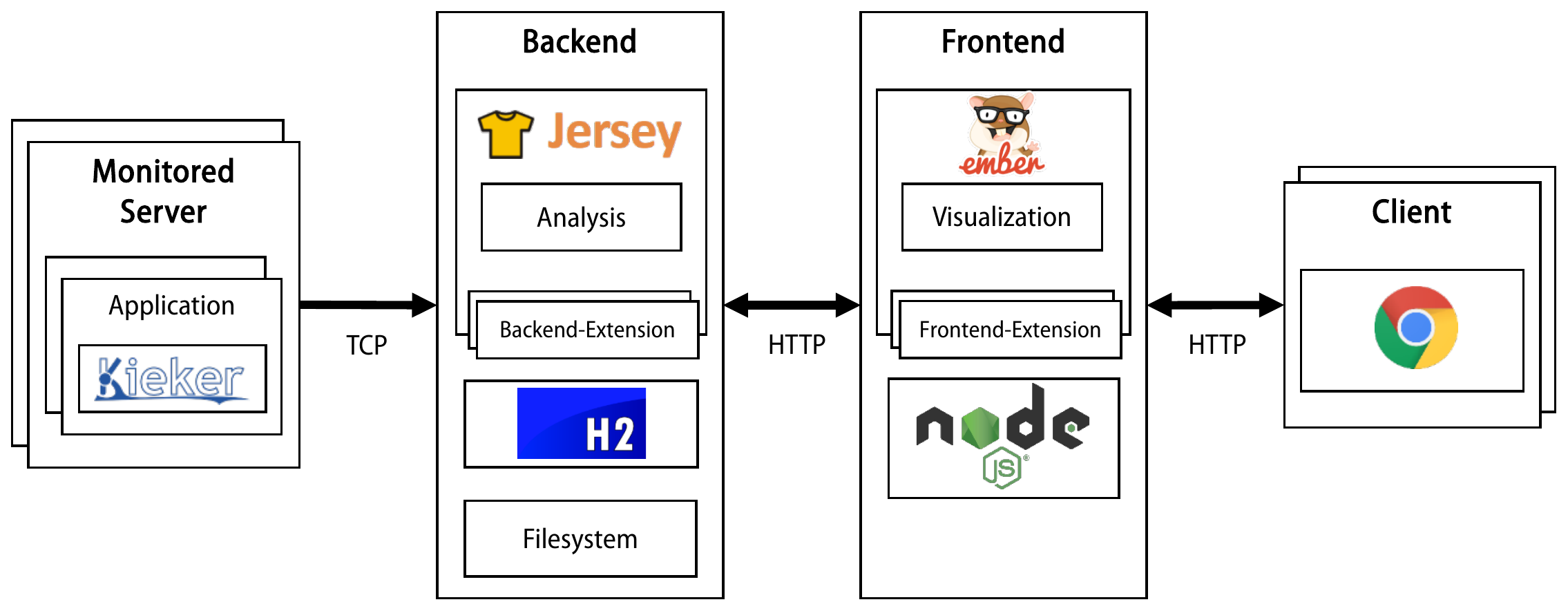}
	\caption{Architectural overview and software stack of the modularized~\projectName.}
	\label{fig:architecture-stack-after}
\end{figure*}

We realized a proof-of-concept implementation and split our project as planned into two separate projects -- a backend project based on \textit{Jersey}, and a frontend project employing the JS framework \textit{Ember}.
Both frameworks have a large and active community and offer sufficient documentation, which is important for new developers.
As shown in~\Cref{fig:architecture-stack-after}, we strive for an easily maintainable, extensible, and plug-in-oriented microservice architecture. 
Since the end of our modularization and modernization process in early 2018, we were able to successfully develop several extensions both for the backend and the frontend. Two of them are described in the following.

\subsubsection{Application Discovery}

Although we employ a monitoring framework, it lacks a user-friendly, automated setup configuration due to its framework characteristics.
Thus, users of~\projectName~experienced problems with instrumenting their applications for monitoring.
In~\cite{ExplorViz:ApplicationDiscovery:2018}, we reported on our application discovery and monitoring management system to circumvent this drawback.
The key concept is to utilize a software agent that simplifies the discovery of running applications within operating systems.
Furthermore, this extension properly configures and manages the monitoring framework.
The extension is divided in a frontend extension providing a configuration interface for the user, and a backend extension, which applies this configuration to the respective software agent lying on a software system.

Finally, we were able to conduct a first pilot study to evaluate the usability of our approach with respect to an easy-to-use application monitoring.
The improvement regarding the usability of the monitoring procedure of this extension was a great success.
Thus, we recommend this extension for every user of~\projectName.

\subsubsection{Virtual Reality Support}

An established way to understand the complexity of a software system is to employ visualizations of software landscapes.
However, with the help of visualization alone, exploring unknown software is still a potentially challenging and time-consuming task.
For this extension, three students followed a new approach using Virtual Reality~(VR) for exploring software landscapes collaboratively. 
They employed head mounted displays (HTC Vive and Oculus Rift) to allow the collaborative exploration of software in VR. 
They built upon our microservice architecture and employed WebSocket connections to exchange data to achieve modular extensibility and high performance for this real-time user environment. 
As a proof of concept, they conducted a first usability evaluation with 22 probands.
The results of this evaluation revealed a good usability and thus constituted a valuable extension to~\projectName.

\section{Restructured Architecture and new Process}\label{sec-ongoingWork}

\begin{figure*}[bt]
	\centering
	\includegraphics[width=\textwidth]{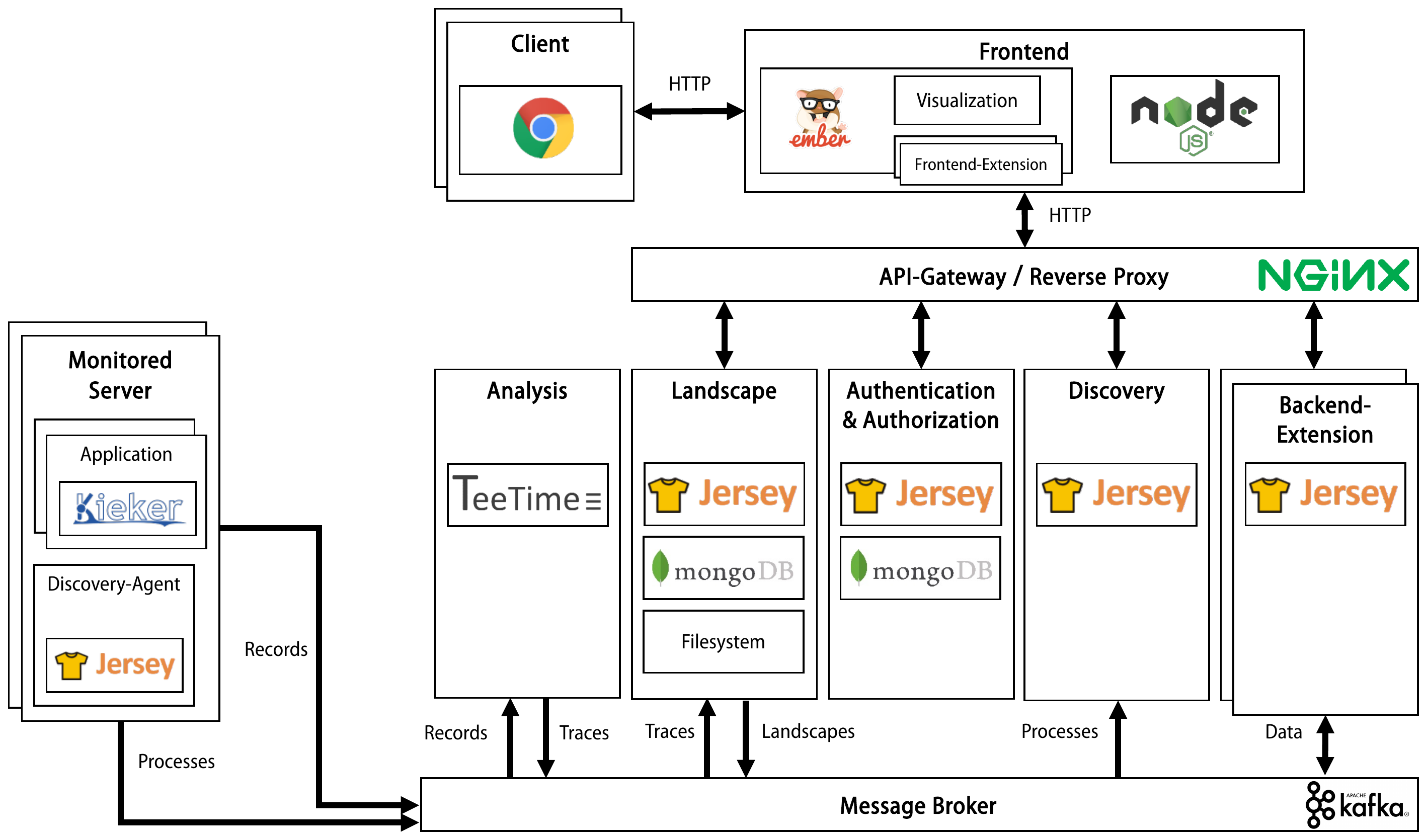}
	\caption{Architectural overview and software stack of the restructured~\projectName.}
	\label{fig:architecture-stack-ongoing}
\end{figure*}
Our modularization approach started by dividing the old monolith into separated frontend and backend projects~\cite{EMLS:2018}.
Since then, we further decomposed our backend into several microservices to address the problems stated in~\Cref{sec-problemStatement}.
The resulting, restructured architecture is illustrated in~\Cref{fig:architecture-stack-ongoing} and the new collaborative development process is described below.
As reported in~\Cref{sec-proofOfConcept}, the new architecture already improved the collaboration with new developers who realized new features as modular extensions.

\subsubsection{Extensibility \& Integrability}
Frontend extensions are based on \textit{Ember's} addon mechanism.
The backend, however, used the package scanning feature of \textit{Jersey} to include extensions.
The result of this procedure was again an unhandy configuration of a monolithic application with high coupling of its modules.
Therefore, we once again restructured the approach for our backend plug-in extensions.
The extensions are now decoupled and represent separated microservices.
As a result, each extension is responsible for its own data persistence and error handling.
Due to the decomposition of the backend, we are left with multiple Uniform Resource Identifiers~(URI).
Furthermore, new extensions will introduce additional endpoints, therefore more URIs again.
To simplify the data exchange handling based on those endpoints, we employ a common approach for microservice-based backends.
The frontend communicates with an API gateway instead of several single servers, thus only a single base Uniform Resource Locator~(URL) with well-defined, multiple URIs.
This gateway, a \textit{Nginx} reverse proxy~\cite{Software:Nginx}, passes requests based on their URI to the respective proxied microservices, e.g., the landscape service.
Furthermore, the gateway acts as a single interface for extensions and offers additional features like caching and load balancing.
Extension developers, who require a backend component, extend the gateway's configuration file, such that their frontend extension can access their complement.
The inter-service communication is now realized with the help of \textit{Apache Kafka}~\cite{Software:Kafka}.
\textit{Kafka} is a distributed streaming platform with fault-tolerance for loosely coupled systems.
The decomposition into several independent microservices and the new inter-service communication approach both facilitate low coupling in our system.

\subsubsection{Code Quality \& Comprehensibility} 
The improvements for code quality and accessibility, which were introduced in our first modularization approach, showed a perceptible impact on contributor's work.
For example, recurring students approved the easier access to~\projectName~and especially the obligatory exchange format \textit{json:api}.
However, we still lacked a common code style in terms of conventions and best practices.
To achieve this and therefore facilitate maintainability, we defined compulsory rule sets for the quality assurance tools \textit{Checkstyle} and \textit{PMD}.
In addition with \textit{SpotBugs}~\cite{Software:Spotbugs}, we impose their usage on contributors for Java code.
For JS, we employ \textit{ESLint}~\cite{Software:ESLint}, i.e., a static analysis linter, with an \textit{Ember} community-driven rule set.
All tools are integrated into our continuous integration pipeline configured in \textit{TravisCI}~\cite{Software:TravisCI}.
\subsection{Software Configuration \& Delivery}
One major problem of \textit{ExplorViz Legacy} was the necessary provision of software configurations for different use cases.
The first iteration of modularization did not entirely solve this problem.
The backend introduced a first approach for an integration of extensions, but their delivery was cumbersome.
Due to the tight coupling at source code level we had to provide the compiled Java files of all extensions for download.
Users had to copy these files to a specific folder in their already deployed \textit{ExplorViz} backend.
Therefore, configuration alterations were troublesome.
With the architecture depicted in~\Cref{fig:architecture-stack-ongoing} we can now provide a jar file for each service with an embedded web server.
This modern approach for Java web applications facilitates delivery and configuration of \textit{ExplorViz}'s backend components.
In the future, we are going to ship ready-to-use Docker images for each part of our software.
The build of these images will be integrated into the continuous integration pipeline.
Users are then able to employ docker-compose files to achieve their custom \textit{ExplorViz} configuration or use a provided docker-compose file that fits their needs.
As a result, we can provide an alternative, easy to use, and exchangeable configuration approach that essentially only requires a single command line instruction.
The frontend requires another approach, since (to the best of our knowledge) it is not possible to install an \textit{Ember} addon inside of a deployed \textit{Ember} application.
We are currently developing a build service for users that ships ready-to-use, pre-built configurations of our frontend.
Users can download and deploy these packages.
Alternatively, these configurations will also be usable as Docker containers.

\section{Related Work}\label{sec-relatedWork}
In the area of software engineering, there are many papers that perform a software modernization in
other contexts. %
Thus, we restrict our related work to approaches, which focus on the modernization of monolithic applications towards a microservice architecture.
\cite{Carrasco:2018} present a survey of architectural smells during the modernization towards a microservice architecture.
They identified nine common pitfalls in terms of bad smells and provided potential solutions for them.
\projectLegacyName\ was also covered by this survey and categorized by the \enquote{Single DevOps toolchain} pitfall.
This pitfall concerns the usage of a single toolchain for all microservices.
Fortunately, we addressed this pitfall since their observation during their survey by employing independent toolchains by means of pipelines within our continuous integration system for the backend and frontend microservices.
\cite{HolgerMicroservices:2018} present a migration process to decompose an existing software system into several microservices.
Additionally, they report from their gained experiences towards applying their presented approach in a legacy modernization project.
Although their modernization drivers and goals are similar to our procedure, their approach features a more abstract point of view on the modernization process.
Furthermore, they focus on programming language modernization and transaction systems.
In~\cite{OttoMicroservices:2017}, the authors present an industrial case study concerning the evolution of a long-living software system, namely a large e-commerce application.
The addressed monolithic legacy software system was replaced by a microservice-based system.
Compared to our approach, this system was completely re-build without retaining code from the (commercial) legacy software system. 
Our focus is to facilitate the collaborative development of open source software and also addresses the development process.  
We are further planning to develop our pipeline towards continuous delivery for all microservices mentioned in~\Cref{sec-ongoingWork} to minimize the release cycles and offer development snapshots.
\section{Conclusion}\label{sec-conclusions}
In this paper, we report on our modularization and modernization process of the open source research software~\projectName, moving from a monolithic architecture towards a microservice architecture with the primary goal to ease the collaborative development, especially with students.
We describe technical and development process related drawbacks of our initial project state until 2016 in~\projectLegacyName~and illustrate our modularization process and architecture.
The process included not only a decomposition of our web-based application into several components, but also technical modernization of applied frameworks and libraries.
Driven by the goal to easily extend our project in the future and facilitate a contribution by inexperienced collaborators, we offer a plug-in extension mechanism for our core project, both for backend and frontend.
We realized our modularization process and architecture in terms of a proof-of-concept implementation and evaluated it afterwards by the development of several extensions of~\projectName.
However, the modularization process is not fully completed, as yet.
We are still improving the project in order to achieve a fully decoupled microservice architecture, consisting of a set of self-contained systems and well-defined interfaces in-between.
In the future, we are planning to evaluate our finalized project, especially in terms of developer collaboration.
Additionally, we plan to move from our continuous-integration pipeline towards a continuous-delivery environment.
Thus, we expect to decrease the interval between two releases and allow users to try out new versions, even development snapshots, as soon as possible.
Furthermore, we plan to use architecture recovery tools like~\cite{MicroART:2017} for refactoring or documentation purposes in upcoming versions of \projectName.

\printbibliography

\end{document}